\documentclass[preprint,showpacs,preprintnumbers,amsmath,amssymb,showkeys,prl]{revtex4-1}

\usepackage{graphicx}
\usepackage{dcolumn}
\usepackage{bm}
\usepackage{epsfig}
\usepackage{amsmath}

\usepackage{fixltx2e}
\begin{document}
\title{Analysis of Anchor-Size Effects on Pinned Scroll Waves and\\ Measurement of Filament Rigidity}
\author{Elias Nakouzi}
\author{Zulma A. Jim\'{e}nez}
\author{Vadim N. Biktashev$^\dag$}
\author{Oliver Steinbock}
\thanks{Corresponding author}
\email{steinbck@chem.fsu.edu\\}
\affiliation{Florida State University, Department of Chemistry and Biochemistry, Tallahassee, FL 32306-4390}
\affiliation{$^\dag$University of Exeter, Department of Mathematical Sciences, Exeter, UK}

\begin{abstract}Inert, spherical heterogeneities can pin three-dimensional scroll waves in the excitable Belousov-Zhabotinsky reaction. Three pinning sites cause initially circular rotation backbones to approach equilateral triangles. The resulting stationary shapes show convex deviations that increase with decreasing anchor radii. This dependence is interpreted as a transition between filament termination at large surfaces and true, local pinning of a continuous curve. The shapes of the filament segments are described by a hyperbolic cosine function which is predicted by kinematic theory that considers filament tension and rigidity. The latter value is measured as  (1.0 $\pm$ 0.7)$\times$$10^{-6}$ cm$^4$/s.
\end{abstract}
\maketitle

\section{INTRODUCTION}

In systems far-from-equilibrium, macroscopic patterns can emerge from processes at the molecular level. These selforganized structures show fundamental universalities across a wide range of disciplines and applications such as type II superconductors \cite{Gurevich05}, neural networks \cite{Huang2004}, geochemical systems \cite{Ortoleva84}, and reaction-diffusion (RD) media \cite{DeKepper08}. Frequently studied examples are two-dimensional spiral waves in excitable and oscillatory RD systems. These wave patterns rotate around a zero-dimensional phase singularity \cite{Muller06}. Their tip describes system-specific trajectories \cite{Engel2002} which, in the simplest case, are circles with radii much smaller than the wavelength of the spiral. These rotors exist in chemical and biological systems such as catalytic surface reactions \cite{Rotermund2012} and giant honey bees defending their nests against hornets \cite{Kastberger2008}. Furthermore, spiral waves have been linked to medical phenomena such as contractions of the uterus during childbirth \cite{Holden2013} and life-threatening cardiac arrhythmias \cite{Davidenko92}. Many of these biological processes occur in sufficiently thick tissue to require a spatially three-dimensional description. Under such conditions, spirals extend to more complex rotors called scroll waves \cite{Gray95}.

A scroll wave can be viewed as a continuum of stacked spirals rotating around a one-dimensional phase singularity \cite{Winfree88}. This curve is called the filament and organizes the surrounding wave field. The motion of the filament is controlled by its own curvature, associated phase gradients ("twist") of the vortex, as well as other factors. Examples include self-shrinking, translating, and chaotic motion. In simple cases the motion of the filament can be described by the kinematic equation
%
\begin{equation}
\frac{d{\bf{s}}}{dt} = (\alpha{\bf{\hat{N}}} + \beta {\bf{\hat{B}}}) \kappa,
\end{equation}
where $\bf{s}$ is a position vector (pointing at the filament) with the corresponding normal ($\bf{\hat{N}}$) and binormal ($\bf{\hat{B}}$) unit vectors, $t$ is time, and $\kappa$ is the local filament curvature. The constants $\alpha$ and $\beta$ are system-specific parameters~\cite{Keener88,Biktashev94}. Several studies have modeled specific cases using this equation e.g. \cite{Keener92,Barkley02}.

For a filament tension of $\alpha$ $\textgreater$ 0, the filament contracts and vanishes in finite time. This case is prevalent among most experimental systems \cite{Vinson97,Ost2006}. Negative values of $\alpha$ cause filament expansion and lead to scroll wave turbulence \cite{Biktashev98,Henry02,Ost2007}. The translational drift coefficient $\beta$ controls the motion in the out-of-plane direction and equals zero in systems where the activator and inhibitor species have the same diffusion coefficient \cite{Keener88,Alonso04}.

The pinning of vortex waves to unexcitable heterogeneities \cite{Pertsov00,Ost2012b} is of importance not only for fundamental reasons but also due to its potential relevance to cardiac arrythmia. For instance, tachycardia is caused by rotating waves of electrical activity. Recent experimental results suggest that these reentrant waves can become pinned to heterogeneities such as remodeled myocardium \cite{RemodMyo}. Furthermore, tachycardia might develop into a turbulent state (ventricular fibrillation) which is a leading cause of sudden cardiac death \cite{Tusscher07}. The influence of pinning sites on vortices in this turbulent state is entirely unknown and also our understanding of scroll wave pinning in non-turbulent cases is poorly developed.

Recently developed experimental procedures allow the deliberate pinning of 3D vortices in chemical RD media and have opened up a wide range of opportunities for controlled studies. Examples include investigations of scroll waves pinned to inert obstacles such as tori, double-tori, spheres, and cylinders \cite{Ost2009,Ost2010b,Ost2011,Ost2012b}. These experiments have revealed that pinning qualitatively alters the evolution of the filament and often results in life-time enhancement or complete stabilization of vortices that in the absence of pinning sites, would rapidly shrink and annihilate. For instance, Jim\'{e}nez and Steinbock \cite{Ost2010b} reported that filament loops pinned to three and four spherical heterogeneities converge to nearly polygonal filaments while two pinning sites either fail to stabilize the vortex or create lens-shaped filaments that are stabilized by short-range filament repulsion. In addition, it was shown that filaments have the tendency to self-wrap around thin heterogeneities \cite{Ost2012b} suggesting that stable, pinned scroll waves should be a common feature in heterogeneous systems.

Filament pinning is governed by two basic rules. Firstly, the total topological charge over any (external or internal) closed surface must equal zero \cite{Winfree83,Pertsov00}. This condition assigns an individual topological charge to the end point of each filament. For an $n$-armed vortex this charge has an absolute value of $n$ and a sign that reflects the sense of rotation. A simple example is the circular filament of a one-armed scroll ring pinned to spherical heterogeneities (see Fig.~1(a)). The filament loop touches each sphere twice and the corresponding charges are +1 and -1 because (as viewed from the sphere's interior) the rotation sense of the local spirals is different. A similar example is a straight filament spanning from one external wall of a box-like system to another. Secondly, a filament ending on a no-flux boundary must be oriented in normal direction to the surface \cite{Meron92}. This condition might be violated only in singular events such as a collision of a filament with a wall but will reestablish itself very quickly.

In this Article, we utilize the Belousov-Zhabotinsky (BZ) reaction as a convenient experimental model system for the study of scroll wave pinning. Traveling waves in this system are driven by the autocatalytic production and diffusion of bromous acid. Our experiments reveal a seemingly small effect that exposes an unexpected difference between filament pinning to small objects and filament termination at large objects. These two limiting cases are illustrated in Figs.~1(b,c), respectively and address the question in how far the surrounding wave fields enforce a smooth, "kink-free" transition of the filament line through the inert and impermeable heterogeneity. In addition, we show that the observed effects cannot be explained in the framework of Eq. (1) but require a higher-order term that we interpret as filament rigidity \cite{Biktashev12}. We also report the first measurement of this quantity.

\section{EXPERIMENTAL METHODS}

The BZ system consists of a bottom gel layer and a top aqueous layer, each of thickness 0.48~cm. The reactant concentrations are identical in the two layers: [NaBrO\textsubscript{3}] = 0.04~mol/L, [malonic acid]= 0.04~mol/L, [H\textsubscript{2}SO\textsubscript{4}] = 0.16~mol/L, and [Fe(phen)\textsubscript{3}SO\textsubscript{4}] = 0.5~mmol/L. At the present gel composition (0.80\% agar w/v), also all diffusion coefficients are expected to be identical throughout the system. For these reaction conditions, the filament tension is found to be $\alpha = (1.4 \pm 0.2) \times 10^{-5}$ cm\textsuperscript{2}/s \cite{Ost2012}, which implies that free scroll rings collapse. Additionally, free filament motion in binormal direction is not observed and we can hence assume that $\beta \approx 0$.

After preparation of the BZ solutions, the pre-gel component is poured into a Petri dish, and three spherical beads are embedded halfway through the forming gel surface. These spheres serve as the inert and impermeable heterogeneities capable of anchoring the filaments of the scroll waves. The aqueous layer is then poured on top and the tip of a silver wire is transiently positioned at the center of the three-bead triangle thus instigating an expanding spherical wave. This excitation wave is nucleated by the local formation of AgBr which decreases the concentration of the inhibitory bromide ion and prompts an oxidation wave. The system is then manually swirled to homogenize the top liquid layer while the half-spherical wave expands in the gel below. As the wave approaches the bead heterogeneities, the swirling action is stopped and the fluid is allowed to come to rest. A flat glass plate is placed on top of the liquid plate to prevent the inflow of atmospheric oxygen, the release of gaseous bromine, and evaporation. Thereafter, the upper rim of the half-sphere curls inward and nucleates a scroll ring. Due to the nature of this procedure, its filament is coplanar, in very close vicinity to the gel-liquid interface, and pinned to the three beads.

The experimental system is monitored from above using a CCD camera equipped with a blue dichroic filter. Concentration waves can be observed for more than six hours after which the depletion of reactants extinguishes the patterns. The formation and growth of reaction-generated CO\textsubscript{2} bubbles does not affect our experiments for at least the first four hours of reaction. All experiments take place at a constant temperature of $ 21.5\,^{\circ}{\rm C}$. Subsequent analyses are performed using in-house MATLAB scripts.

\section{RESULTS}

Figure~2 shows still images from two experiments that differ only in the radius of the employed beads and the resulting wave patterns. Image contrast stems from the varying ratio between the chemically reduced and oxidized form the catalyst (ferroin/ferriin). Accordingly bright and dark regions can be interpreted as excited and excitable areas, respectively. However, this situation is complicated by (unresolved) vertical variations along the depth of the three-dimensional reaction medium. Despite this limitation, the filament loop can be readily detected from image sequences because, due to the wave rotation around the filament, it sequentially emits waves in the outward and inward direction. As such, we can extract the filament coordinates by locating the set of pixels which undergo minimal contrast change for the period of one full rotation. This method for filament detection was pioneered by Vinson et al. \cite{Vinson97}.

Figure~2 shows the result of this analysis as superposed bright (cyan) curves. At the given reaction time of 120~min (which corresponds to approximately 15 rotation periods) the initially circular filament loops have reached a stationary, polygon-like shape. We reemphasize that in the absence of pinning sites, the filament would remain a circle, self-shrink, and vanish. This collapse clearly does not occur. In the following, we characterize the relaxation dynamics into the stabilized, pinned state in terms of the distance $d$ between the center of a given bead and the opposite filament segment. The distance is measured along the height of the equilateral triangle defined by the three pinning sites (see inset in Fig.~3). Two representative examples for the temporal evolution of $d$ are shown in Fig.~3. The data are well described by compressed exponentials of the form

\begin{equation}
d(t) = d_{ss} + (d_{0}-d_{ss})e^{(-t/\tau)^b}.
\end{equation}

\noindent The values $d_0$ and $d_{ss}$ denote the initial loop diameter and the asymptotic distance, respectively. Analysis of various data sets yields an average $b$ value of 2.35, which is $68 \%$ larger than the earlier reported value ($b$~= 1.4) for filaments pinned to two beads \cite{Ost2012}. The time constant $\tau$ varies between 45 to 85 minutes. The high values within this range are typically found for larger filament loops. Note that the observed dynamics are qualitatively different from the contraction of free circular filaments for which the diameter is well approximated by the simple square root law $d(t)^2 = d_0^2 - 8\alpha t$.

As suggested by the examples in Fig.~2, the shape of the stationary filament as well as the corresponding bead-to-filament distance $d_{ss}$ depend on the bead radius $R$. The results of systematic measurements are shown in Fig.~4 for a constant inter-bead distance of $\Delta$~= 7~mm. The steady-state value $d_{ss}$ decreases with increasing values of $R$. Furthermore, the data fall within two simple geometric limits. Firstly, considering filament repulsion and the system's tendency to establish smooth wave patterns, it is unlikely that the tangential vectors of the terminating filament pair can form an angle above 180~degrees. Accordingly, the circle defined by the three bead centers constitutes the upper limit of $d_{ss} \geq 2 \Delta / \sqrt{3}$. Secondly, only filament attraction could establish a filament angle below 60~degrees but would also cause the detachment of the filament from the pinning bead. Hence, the equilateral triangle with corner points in the bead centers creates the lower limit of $d_{ss} \leq \sqrt{3} \Delta / 2$. In Fig.~4 both $d_{ss}$ limits are plotted as dashed lines. We find that our data are closer to the triangular than the circular limit but nonetheless span about 40~$\%$ of the possible range. Unfortunately, we are not able to pin vortex loops for bead radii below 0.75~mm, possibly because of limitations created by the size of the vortices' natural rotation core. The continuous lines in Figs. 4 and 5 relate to a theoretical description that is discussed later.

The experimental data shown in Fig.~4 provide valuable information regarding our initial question whether filament termination at large surfaces is qualitatively different from filament pinning to small heterogeneities. The former case corresponds to experiments with large beads and, because filament interaction is absent, to the limiting case of triangular filaments. The latter case corresponds to very small beads and the circular limit. Given the negative slope of the data in Fig.~4, we find that our data strongly support the hypothesis of qualitative differences between filament termination and pinning. One can speculate that the transition from filament pinning to termination occurs if the circumference of the heterogeneity is comparable to the wavelength of the vortex structure. For our BZ system, the wavelength is $\lambda$~= 4.85~mm which suggests a transition at $R \sim $ 1~mm. Our data are consistent with this value.

The size and shape of the stationary, pinned filament depend not only on the bead radius but also on the inter-bead distance $\Delta$. We therefore performed experiments with vortices pinned to beads arranged at the corners of equilateral triangles with side lengths between 5~mm and 12~mm. The bead radius is kept constant at $R$~= 1.0~mm. Figure~5(a) shows the resulting data which reveal that $d_{ss}$ increases with increasing values of $\Delta$. The dashed lines again represent the limiting cases of perfect circles and triangles. To obtain a better understanding of these results, we remove the trivial linear scaling of $d_{ss}(\Delta)$ and find for increasing values of the bead distance a smooth transition from the triangular limit toward the circular limit (Fig.~5(b)). Experiments with even larger values of $\Delta$ have not yielded reliable data yet because, among other complications, rogue waves and bulk oscillations tend to interfere with the preparation of these large scroll rings.

Additional information on the stationary vortex states can be obtained by analyzing the shape of the filament segments that extend from one bead to another. In most experiments, the deviations between the three individual segments are small and the results shown in the following are the average shape of the stationary filament segment as measured from different experiments and different sides but for identical experimental conditions. Figure~6 shows filament shapes for two different values of $R$. The $s$ axis extends along the line connecting the two bead centers and is zero halfway between those centers. The ordinate is a generalized form of the earlier $d$ variable and measures the distance of the filament from the line that is parallel to the $s$ axis and passes through the third bead center. We find that the filament shape $d(s)$ is in excellent agreement with the hyperbolic cosine function

\begin{equation}
d(t) = d_{ss} - k \cosh{q s}
\end{equation}

\noindent as demonstrated by the solid and dashed curves in Fig.~6. Notice that the values of $d_{ss}$ discussed above (e.g. in the context of Figs.~4,~5) are obtained from direct measurements of the filament position and not from fits to Eq.~(3).


\vspace{0.5cm}

To gain a better understanding of these experimental results, we performed a semi-phenomenological analysis of the system. We suggest that the filament shape is determined by the delicate interplay between the filament tension ($\alpha$) and the filament rigidity ($\epsilon$). Accordingly the equation

\begin{equation}
\partial_t \vec{s} = \alpha \partial_\sigma^2 \vec{s} - \epsilon (\partial_\sigma^4 \vec{s})_\perp
= 0
\end{equation}

\noindent describes the equilibrium of the planar filament $\vec{s}(\sigma)$ with $\sigma$ denoting its arclength. This equation is a special case of the asymptotic theory presented in \cite{Biktashev12} for which several terms (not shown) vanish because (i) the diffusion coefficients of the chemical species in our reaction differ only slightly and (ii) the filament curvature is not too high. A more detailed description and analysis of this kinematic model is presented as supplementary material \cite{SuppMat}. This analysis also considers the short-range, repulsive interaction between the filament segments in the vicinity of the beads, which we describe by a simple exponential decay with a decay constant $p$. This assumption is in accord with experimental and numerical studies of spiral and scroll wave interaction \cite{Aranson1991,Bray2003,Ost2012}.

If we assume that the arclength $\sigma$ of the filament segment differs only slightly from the linear space coordinate $s$, Eq.~(4) readily yields the experimentally observed hyperbolic cosine function (Eq.~(3)) for the shape of the stationary filament. Furthermore, it identifies the fitting parameter $q$ as the square root of the ratio between the filament tension and the filament rigidity

\begin{eqnarray}
&d(s) = d_{ss} - \dfrac{k}{q^2} \cosh{q s} ,\\
&q = \sqrt{\alpha/\epsilon} ,
\end{eqnarray}

\noindent where $k$ is an integration constant. Since the filament tension for this BZ system is known ($\alpha$=(1.4 $\pm$ 0.2) $\times$ $10^{-5}$ cm$^2$/s, \cite{Ost2012}), Eqs.~(5,6) allow us to measure the system-specific filament rigidity $\epsilon$ from the shape of the stationary filament. The results are summarized in Table~1. The average value of $\epsilon$ is (1.0 $\pm$ 0.7) $\times$ $10^{-6}$ cm$^4$/s. We believe this to be the first measurement of filament rigidity in any excitable or oscillatory system.

Notice that the hyperbolic cosine function can be difficult to distinguish from a quadratic polynomial. Resolving fourth order terms, however, is necessary for the desired measurement of $q$ and $\epsilon$ because Eq.~(5) implies that the second-order term has no $q$-dependence ($d(s) = const - \tfrac{k}{2} s^2 + O(s^4)$). Although both functions involve three fitting parameters, the lowest root mean square deviations are generally obtained for fits with the hyperbolic cosine function. For example, the upper data set in Fig.~6 (closed, blue circles) yields root mean square errors of 9.3~$\mu$m and 10.9~$\mu$m for the cosh and quadratic fits, respectively. For the lower data set (open, red circles), these numbers are 4.6~$\mu$m and 5.2~$\mu$m, respectively. Our model predicts that significantly larger values of $\Delta$ (twice and more) cause a transition from a rigidity-dominated to a tension-controlled filament with the latter showing more pronounced deviations from a parabola. However, the execution of such experiments is, as mentioned above, fraught with technical difficulties.

Our theoretical analysis also describes the experimentally observed dependencies of stationary filament distance. As detailed in the Supplemental Material \cite{SuppMat}, we obtain

\begin{equation}
d_{ss} (\Delta,R) = \frac{\sqrt{3}}{2} \Delta + (\frac{1}{p C}-\frac{\pi}{6 q S}) (R q S + C - 1) ,
\end{equation}

\noindent where $S = \sinh{q(\Delta/2-R)}$ and $C = \cosh{q(\Delta/2-R)}$. Equation~(7) can be readily compared to the measurement results in Figs.~4 and 5. For this purpose, we use the average $q$ value of 3.74~cm$^{-1}$ which corresponds to the measured filament tension and rigidity (Eq. (6)) and the appropriate constant bead distance (for Fig.~4) and constant bead radius (for Fig.~5).  Least square fitting then yields the continuous lines in the latter figures, which are in very good agreement with the experimental data. The fits in Figs.~4 and 5 are carried out separately for each data set and yield $p$ values of 3.41~cm$^{-1}$ and 3.03~cm$^{-1}$, respectively.

We also investigated the angle $\phi_0$ between the tangent to the filament at the bead and the $s$-axis. While the data are not fully conclusive (Table~1), we can tentatively identify some trends. Firstly our measurements show that there is only a mild dependence of $\phi_0$ on the bead radius $R$. Secondly, the experimental data suggest that $\phi_0$ increases with increasing values of the distance $\Delta$ to saturate at approximately 50$^\circ$. This saturation behavior seems reasonable if one considers that the central portion of long filaments (large $\Delta$) is increasingly flat. A theoretical description of the angle $\phi_0$ is presented in the Supplemental Material \cite{SuppMat} and is in agreement with the experimentally observed trends.

All of the above experiments and analyses are carried out for scroll filaments pinned to spherical objects located at the corners of equilateral triangles. Figure~7 shows an example of a pinning experiment in which the beads form an isosceles triangle in which the unique angle measures 40\textsuperscript{$\circ$}. The still frames are taken at four different times and the corresponding filament coordinates are superposed in bright cyan color. The filament loop in Figs.~7(a,b) is pinned to all three beads and its curve-shrinking dynamics are similar to the behavior shown in Fig.~2. However, instead of establishing a stationary state, the filament unpins from the lower bead to create a strongly curved segment that quickly withdraws toward the upper bead pair [Fig. 7(c)]. The resulting filament loop is not stable and annihilates as shown in Fig.~7(d). The unpinning from the bottom bead is a direct consequence of the small angle between the two lower filament segments. This process shares similarities with electric-field-induced unpinning events that were recently reported by Jim\'{e}nez and Steinbock \cite{Ost2013}. In addition, it is reminiscent of filament loops detaching from planar no-flux boundaries and hence also related to filament reconnections \cite{recon}. Lastly, we note that a recent study of scroll rings pinned to two spherical heterogeneities suggest that detached filament loops, like the one shown in Fig.~7(c), could generate stable, lens-shaped structures if the inter-bead distance $\Delta$ is sufficiently large. More experiments are needed to test this prediction.

\section{CONCLUSIONS}

Our experiments provide strong evidence that filaments ending at no-flux boundaries show different behavior depending on the size of the impermeable heterogeneity. In the simplest case, a filament terminates at a planar, external wall or other large structure. In this situation, wave rotation around the obstacle is irrelevant. However, if the circumference of the heterogeneity is comparable to or smaller than the pattern wavelength, rotation must be considered and the overall filament will feature shape variations that aim to reduce gradients of the surrounding wave field. The latter force can be also interpreted in terms of a repulsive interaction between the filament segments that end on the same surface.

The global filament shapes resulting from this local pinning process cannot be explained solely by the contractive motion of the underlying curvature flow (Eq.~(1)) but rather reveal a higher-order phenomenon that we refer to as filament rigidity. For the specific situation of three pinning sites located at the corners of an equilateral triangle, the individual filament segments are described by hyperbolic cosine functions that deviate only slightly from simple parabolas. We note that this outcome is reminiscent not only of the shape but also of the history of the catenary ("chain") curve which Galileo described as an approximate parabola \cite{Fahie1902}.

We propose that filament rigidity is also crucial for explaining stationary and/or long-lived states of filaments pinned to thin, cylindrical heterogeneities. Such cases were recently reported for three-dimensional BZ systems \cite{Ost2012b} in which filament loops were attached to long glass rods. From the end points of these rods, the rotation backbone of the vortex extended as a free, C-shaped filament segment that despite its curvature remained essentially stationary. Similarly to our new results, this equilibrium state was likely the result of the antagonistic interplay between filament tension and rigidity.

We believe that future studies should aim to characterize the exact boundary conditions for pinned filaments as the current lack of this information severely hampers more detailed analyses of the experimental observed dependencies. Other challenges include the study of scroll wave pinning to non-inert heterogeneities such as regions with decreased excitability and/or diffusion coefficients. Filaments pinned to such "soft" anchors can be expected to unpin more readily than filaments attached to inert and impermeable pinning sites. We emphasize that both types of anchor regions are relevant to excitable and oscillatory RD systems in biology where heterogeneities result from numerous sources including variations in cell density, cell type, and gap junctions as well as anatomical features such as blood vessels.

\vspace{0.5cm}
{\centerline{\bf{ACKNOWLEDGEMENT}}\setlength{\parskip}{4pt plus 1pt minus 1pt}\par
This material is based upon work supported by the National Science Foundation under Grant No. 1213259.

\newpage


\newpage
\vspace{0.5cm}
{\small TABLE I: Filament rigidity $\epsilon$ and angle $\phi_0$ at the bead measured for different values of the bead radius $R$ and the bead distance $\Delta$.}
\begin{center}
  \begin{tabular}{ c c c c  }
    \hline \hline
     $R$  &  $\Delta$  &  $\epsilon$  &  $\phi_0$ \\
  (mm) &  (mm) & $(\times 10^{-6} cm^4/s)$  &  (degrees) \\ \hline \hline
    0.75 & 7 & 0.37 & 33\\
    1 & 7 & 0.36 &32\\
    1.5  & 7 & 0.36 &33      \\
    2 & 7 & 1.29 &23            \\
    1 & 5 & 1.92     &18        \\
    1 & 6 & 0.21   &34        \\
    1  & 8 & 1.14     &28      \\
    1  & 10 & 1.14    &55   \\
    1  & 12 & 2.24    &45 \\
    \hline
\hline
  \end{tabular}
\end{center}

\newpage
\centerline{\includegraphics[width = 7.5cm]{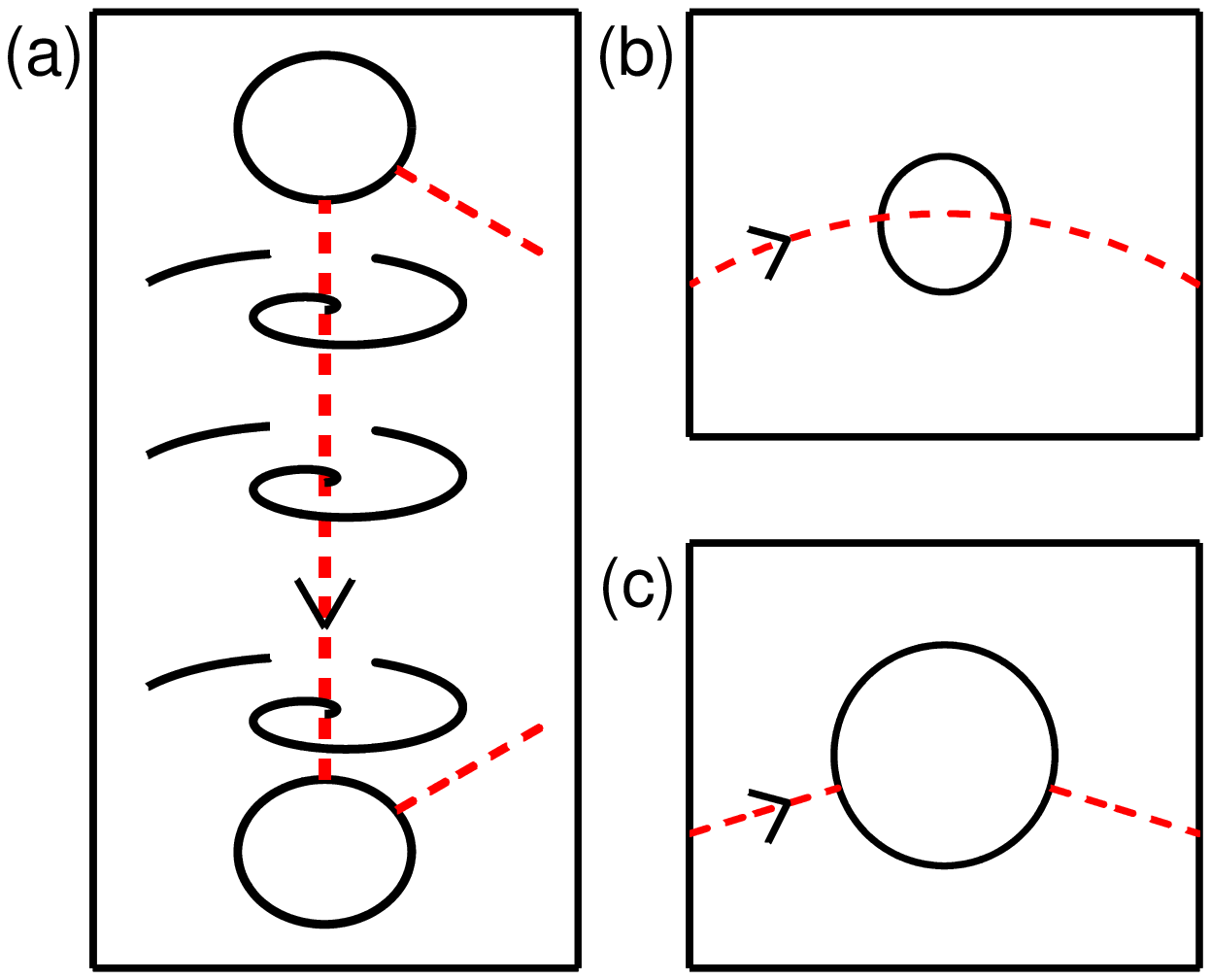}}
\small{ FIG. 1 (Color online) Schematics of a filament (dashed, red line) pinned to two inert spheres (a). The smaller panels show the limiting cases of a locally pinned, smooth filament (b) and a filament cut into two independent segments (c).}

\newpage
\centerline{\includegraphics[trim = 0.05cm 3.8cm 0 1.7cm,  clip, width = 8.5cm]{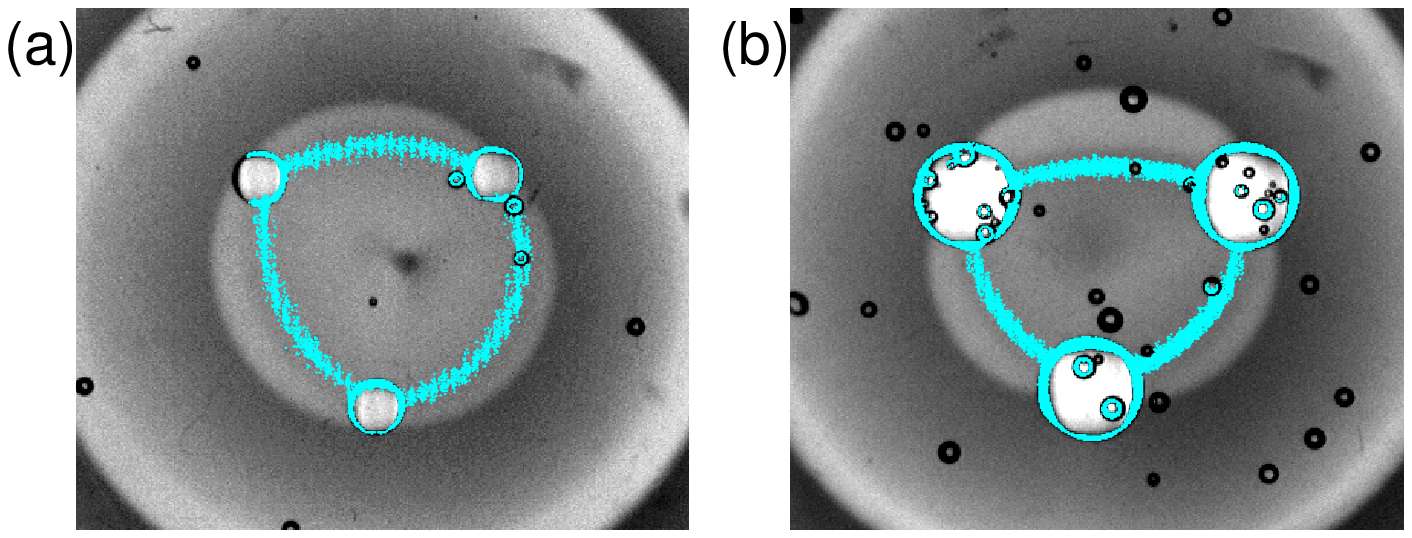}}
{\small FIG. 2 (Color online) Top-view of scroll rings in 1~cm thick layers of BZ reagent. The superposed filaments (bright, cyan) are pinned to three beads (white) of radius 0.75~mm (a) and 1.5~mm (b). The filaments are stationary and slightly more curved in (a) than in (b). The center-to-center bead distance and image area are 7~mm and (16.5 $\times$ 14)~mm\textsuperscript{2}, respectively. The images are obtained 120~min after the start of the reaction and show some CO\textsubscript{2} bubbles (dark spots).}

\newpage
\centerline{\includegraphics[width = 8cm]{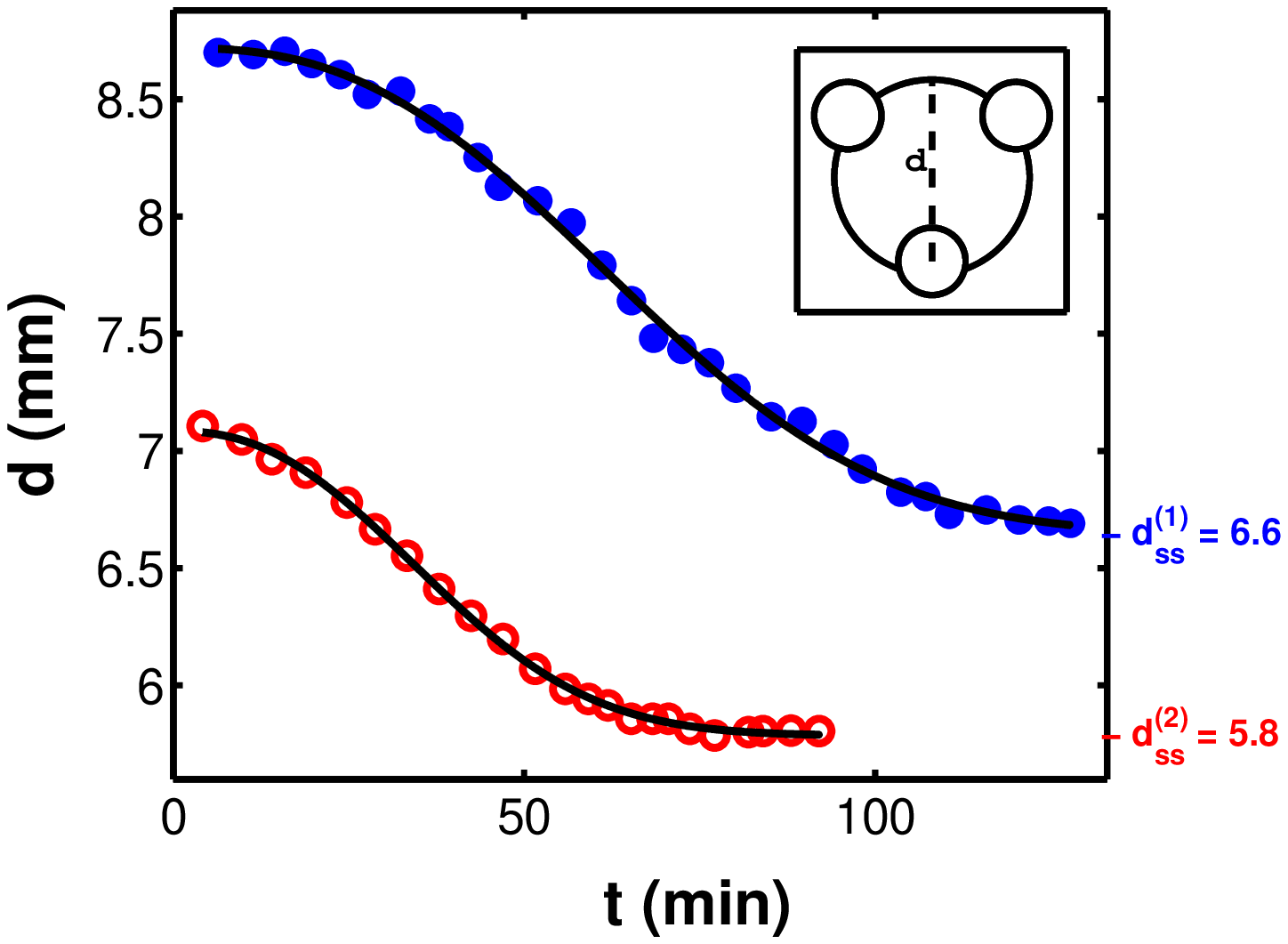}}
{\small FIG. 3 (Color online) Relaxation dynamics of the pinned filament into the stationary state. The distance $d$ is measured from one bead center to the midpoint of the filament fragment opposite to it. This distance is shown schematically in the inset (dashed line). The experimental data sets are obtained for a bead distance of 6 mm (open, red circles) and 7 mm (closed, blue circles). The bead radius in both cases is 1~mm. Solid curves represent best-fit compressed exponentials.}

\newpage
\centerline{\includegraphics[trim = 0 0 3.5cm 0,  clip,width = 8cm]{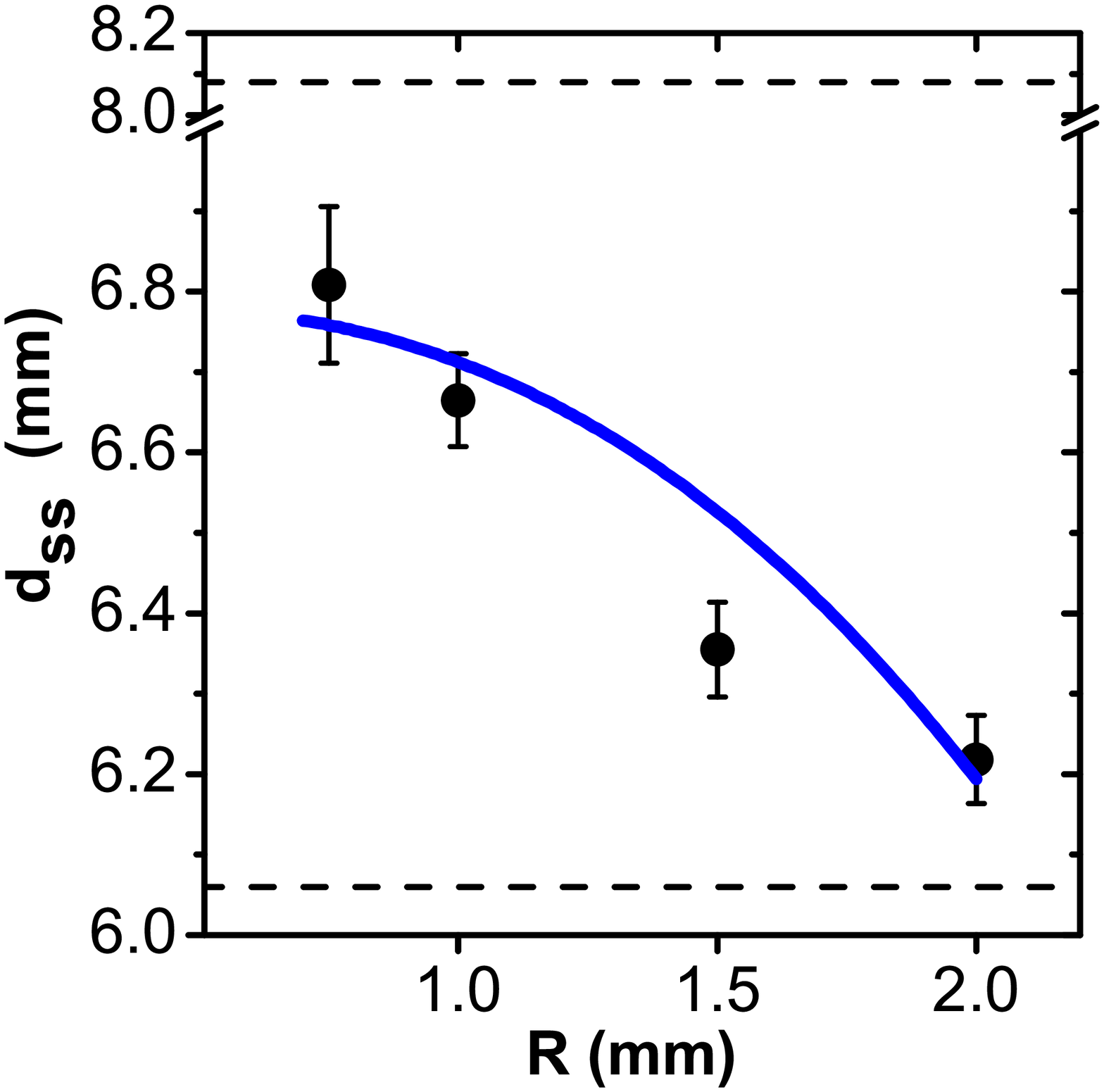}}
{\small FIG. 4. Distance $d_{ss}$ of the stationary filament from the opposite bead as a function of the radius $R$ of the pinning beads. The distance is measured along the central symmetry line of the equilateral bead triangle. The two dashed lines are the geometric limits of a circular (top) and triangular (bottom) filament. With respect to the three bead centers, they equal the diameter of the circumcircle and the height of the bead triangle, respectively. The continuous curve is the best fit of Eq. (7) to the experimental data. The bead distance is kept constant at 7~mm.}

\newpage
\centerline{\includegraphics[width = 8cm]{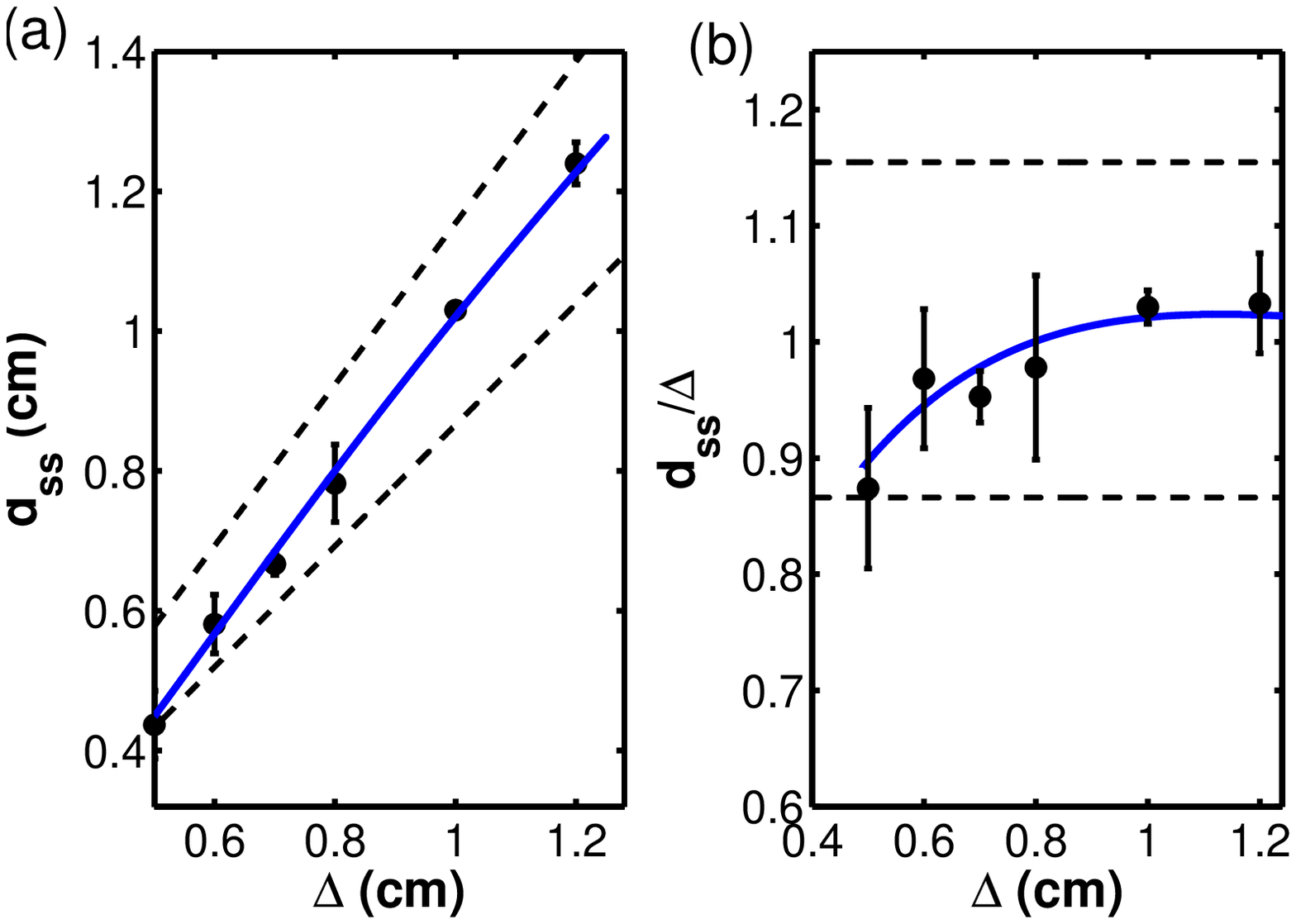}}
{\small FIG. 5. Distance $d_{ss}$ of the stationary filament from the opposite bead as a function of the bead distance $\Delta$ (a). In (b) the same data are shown in terms of the ratio $d_{ss}/\Delta$. As in Fig.~4, the dashed lines correspond to the geometric limits of a circular (top) and triangular (bottom) filament. The continuous curves are the best fit of Eq. (7) to the experimental data. The bead radius is kept constant at 1.0~mm. \vspace{0.5cm}}

\newpage
\centerline{\includegraphics[width = 8cm]{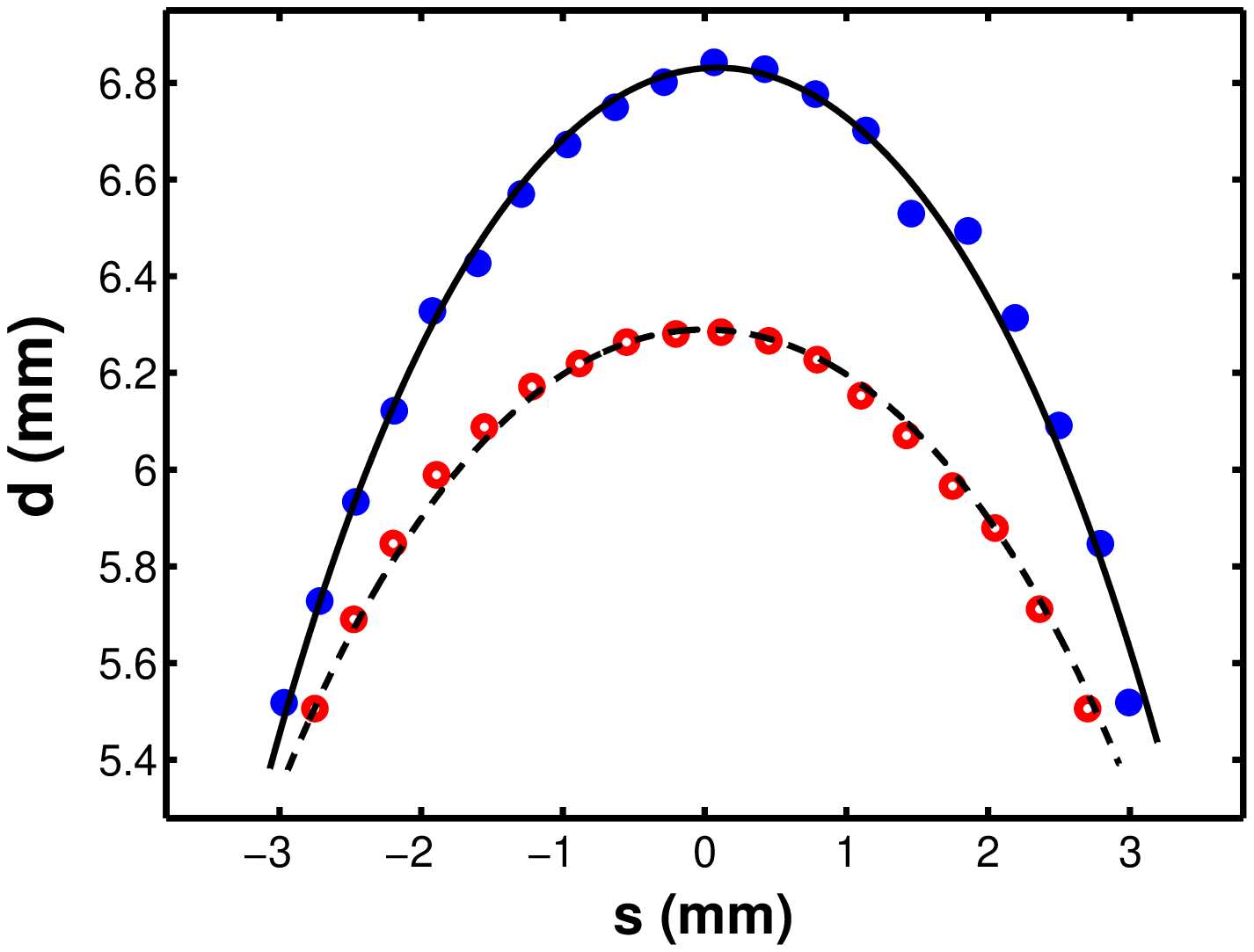}}
{\small FIG. 6 (Color online) Averaged shape of the stationary filament for $R$~= 0.75~mm (open, red circles) and 1.5~mm (closed, blue circles). The dashed and solid curves represent the corresponding hyperbolic cosine fits (Eq.~(3)) which allow the measurement of the filament rigidity $\epsilon$ (see Table~1). For both experiments the bead distance is $\Delta$~= 7~mm.\vspace{0.5cm}}

\newpage
\centerline{\includegraphics[width = 9cm]{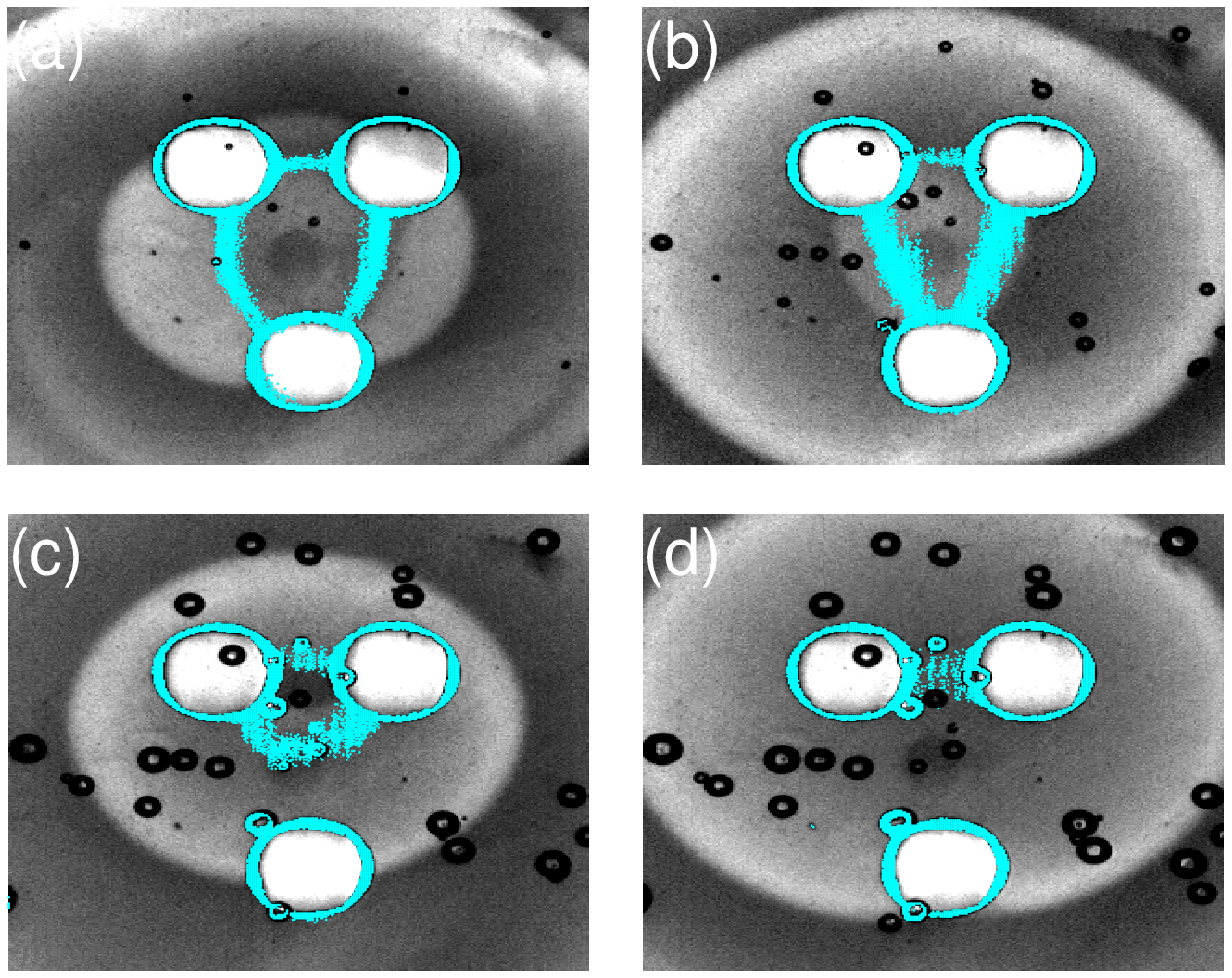}}
{\small FIG. 7 (Color online) Still frames illustrating the evolution of a filament pinned to three beads located on an isosceles triangle. The snapshots are taken at t~= 27~min (a), 67~min (b), 123~min (c), and 140~min (d). The filament unpins at the tight angle and collapses.}

\end{document}


\title{Supplemental Material:\\Analysis of Anchor-Size Effects on Pinned Scroll Waves and\\Measurement of Filament Rigidity}
\author{Elias Nakouzi}
\author{Zulma A. Jim\'{e}nez}
\author{Vadim N. Biktashev$^\dag$}
\author{Oliver Steinbock}
\affiliation{Florida State University, Department of Chemistry and Biochemistry, Tallahassee, FL 32306-4390}
\affiliation{$^\dag$University of Exeter, Department of Mathematical Sciences, Exeter, UK}

\maketitle

\noindent In the main paper, we have derived the shape of the stationary
filament without any detailed assumptions regarding the boundary conditions
at the pinning bead and the nature of the repulsive filament interaction.
In the following, we present a more detailed analysis that, despite the simplifications made, aims to describe additional details including the
dependencies of the central distance $d_{ss}$ between the filament and the
opposing bead center on the bead radius $\r$ and the bead distance $\Side$.
We reemphasize that the following analysis of our experiments is
semi-phenomenological.

We suggest that the filament shape is determined by the delicate interplay
between the filament tension, filament rigidity, and filament repulsion
close to the beads,
\begin{align*}
  \dd_t \R =
  \alpha \dd_\s^2 \R
  - \epsilon \left(\dd_\s^4 \R\right) _\perp
  + \nonsc |\dd_\s^2 \R|^2 \dd_\s^2 \R
  + \left(\F(\R)\right)_\perp
  = 0,                                                       \eqlabel{eom}
\end{align*}
which is an equation of equilibrium of the filament, where the first
three terms are as in the asymptotic theory presented in
\cite{Biktashev12}, with $\tenps=\rigps=\nonps=0$ because the
diffusion coefficients of all the reagents are close to each other,
$\tensc=\alpha$ and $\rigsc=\epsilon$ to agree with
the notations with the main paper,
and
the last term is added to describe phenomenologically the repulsion of
filaments from each other in the vicinity of the bead.

\begin{wrapfigure}[11]{R}{0.4\textwidth}
  \vspace*{-\baselineskip}
  \includegraphics[width=0.4\textwidth]{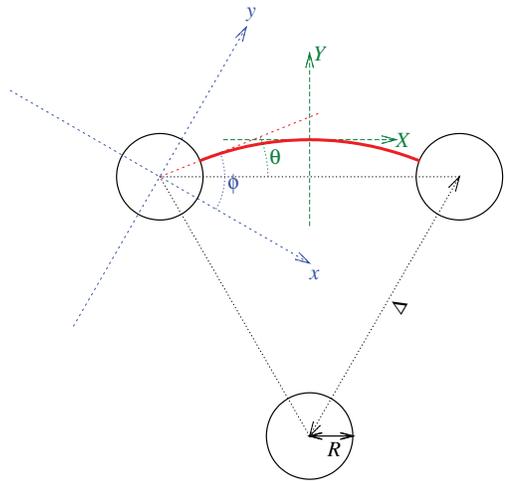}
  \vspace*{-1.5\baselineskip}
  
  \caption{\footnotesize Sizes, coordinates, and angles.}
  \figlabel{angles}
\end{wrapfigure}

We assume that
the repulsion distance is much smaller than the distance between the
beads and the shape of a filament is symmetric so we consider one half
of it, from one bead to its middle point between two beads.  To define
the repulsion "force", we assume that all the filaments lie in one
plane, that two filaments touching the same bead are mirror
reflections of each other, introduce Cartesian coordinates in the
plane with the origin at the center of the bead, $\x$ axis along the
mirror symmetry axis, and focus on the filament with $\y>0$. In these
terms, we assume, following numerical and experimental results in
\cite{Bray2003,Ost2012}, that the
repulsion velocity decreases exponentially with the distance to the
mirror, $|\F|=\amp\exp(-\p\y)$, where $\amp>0$, $\p>0$ are some
constants to be determined empirically, and take the normal component
of the velocity to be further proportional to the cosine of the angle
$\fangle$ of the filament with respect to the mirror line, so e.g. a
filament perpendicular to the mirror line is not affected by its
reflection. Let $\s$ be the arclength coordinate along the filament,
starting from the point it touches the bead, and let $\fangle(\s)$ be the
angle of the tangent vector $\dd_\s\R$ with respect to the $\x$ axis,
so the local filament curvature is $\curv(\s)=\dd_\s\fangle$. Then,
rewriting the above equation, we obtain
\begin{align}
& \curv''= \Ten \curv
         + \Non \curv^3
         + \Amp \, \e^{-\p\y}\cos(\fangle),  \\
& \fangle' = \curv, \\
& \y' = \sin(\fangle),
\end{align}
where $\Ten=\alpha/\epsilon$, $\Non=1+\nonsc/\epsilon$, and
$\Amp=\amp/\epsilon$. The dash designates differentiation with respect to
$\s$ and the last two equations are definitions of $\fangle$ and $\y$
added to complete the system.

The first simplification we make is to assume that the curvature is not
too high throughout the length of the filament, so we can neglect the
$\Non\curv^3$ term.  Further, we assume that the filament does
not change its orientation significantly within the range of the
repulsive velocity. Then the filament curvature profile is described by
the equation
\begin{align}
  & \curv''\approx \Ten \curv
  + \Amp \, \e^{-\p\sin\fangle_0(\r+\s)}\cos(\fangle_0),
\end{align}
the general solution of which is
\begin{align*}
  \curv=C_1\e^{\q\s}+C_2\e^{-\q\s}
  +\frac{\L}{\p^2\sin^2\fangle_0-\q^2}
  \,\e^{-\p\sin(\fangle_0)\s} ,
\end{align*}
where $\q=\Ten^{1/2}$,
$\L=\Amp \cos(\fangle_0) \, \e^{-\p\sin(\fangle_0)\r}$,
and we assume that $\q\ne\p\sin(\fangle_0)$ (an equality here would be a
rare event in any case). The constants $C_1$ and $C_2$ depend
on the boundary conditions.
Ideally, further boundary conditions for
this equation should be obtained from the same asymptotics that were
used to derive the filament equations of motions; these are not
available at present. In our simplified approach, we suppose that both
$\curv(0)$ and $\curv'(0)$ are zero or negligibly small, then
\begin{align}
  C_1 &= \frac{\L}{2\q(\q+\p\sin\fangle_0)}, \nn\\
  C_2 &= \frac{\L}{2\q(\q-\p\sin\fangle_0)} . \eqlabel{C1C2}
\end{align}

The scale of the inner region is $\p^{-1}$ as beyond a few of those
lengths the interaction term is negligible.  The consistency condition
of our approximation is that the change of $\fangle$ throughout the
inner region, which assuming $\sin\fangle_0\propto1$,
$\cos\fangle_0\propto1$, is $\Delta\fangle\propto\Amp/\p^3$, is
small. For $\s$ beyond the inner region, we can neglect the
term proportional to $\exp(-\p\sin\fangle_0\s)$, so the solution is
simply
\begin{align*}
  \curv\approx C_1\e^{\q\s}+C_2\e^{-\q\s} = - \curvm \cosh\left(\q(\s-\sm)\right)
\end{align*}
where $\curvm>0$
is the minimal curvature of the
filament, achieved at the middle point between the beads, and $\sm$ is
the arclength coordinate of the middle point.
With account of \eq{C1C2}, this gives
\begin{align}
  \sin(\fangle_0) = \frac{\q}{\p} \tanh(\q\sm) .             \eqlabel{fangle0}
\end{align}
In particular, this predicts a plateau for $\fangle_0$ at large values
of $\sm$, satisfying
\begin{align}
  \sin(\bar\fangle_0) = \frac{\q}{\p} .                      \eqlabel{fangle0bar}
\end{align}

Notice that the latter relation is in good agreement with the experimental
data presented in Table~1 of the main paper because the measured angles $\fangle_0$
are essentially constant. Even the small decrease of $\fangle_0$ for the largest
bead radius ($R \sim  2\,\mm$) could be due to the $R$-dependent decrease of
$\sm$, consistent with the more accurate formula \eq{fangle0}.
For $\fangle_0\approx50^{\circ}$,
$\q=\left(\alpha/\epsilon\right)^{1/2}$,
$\alpha=1.4\times10^{-5}\,\cm^2/\sec$, and
$\epsilon=10^{-6}\,\cm^4/\sec$, the characteristic length scale of filament repulsion
is then estimated as
$\p^{-1}\approx 0.3\,\cm$. This value appears plausible if we consider the vortex
wavelength of about 0.5~cm (see e.g. Fig.~2 in main paper).

The spatial shape of the filament in the outer region is described
parametrically as
\begin{align*}
  & \X(\ss) = \int\limits_0^{\ss} \cos(\curvm\sinh(\q\ss)/\q) \,\d\ss,
  \\
  & \Y(\ss) = - \int\limits_0^{\ss} \sin(\curvm\sinh(\q\ss)/\q) \,\d\ss,
\end{align*}
where $(\X,\Y)$ are Cartesian coordinates centered at the midpoint with
the $\X$ along the tangent to the filament at that point, and
$\ss=\s-\sm$ is the arclength coordinate measured from the midpoint. These
integrals cannot be evaluated in elementary functions, but using the
assumption of smallness of curvature already made,
in terms of the filament angle $\Fangle$ in the $(X,Y)$
coordinate system, we can approximate
$\sin(\Fangle)\approx\Fangle$, $\cos(\Fangle)\approx1$, leading simply to
\begin{align*}
  \Y(\X) = \frac{\curvm}{\q^2} \left(1-\cosh(\q\X)\right).
\end{align*}
In the same approximation we have
\begin{align*}
  \fangle=\pi/6+\Fangle\approx\frac{\pi}{6}-\curvm\sinh(\q\X)/\q
\end{align*}
where the $\pi/6$ is due to the difference of orientation between $(\x,\y)$
and $(\X,\Y)$ frames (see~\fig{angles}).
Combining these results with the previously obtained~\eq{fangle0}, we
obtain, after elementary transformations, the following
expression for the midpoint curvature
\begin{align*}
  \curvm = \frac{\q^2}{\p\cosh(\q\sm)} - \frac{\pi\q}{6
    \sinh(\q\sm)} .
\end{align*}
We estimate the the distance $\Distance$ from the center of a bead
to the middle of the contraposed filament as $\Distance =
\Side\cos(\pi/6) + \r\sin(\Fangle_0) + |\Y(-\sm)|$ (see~\fig{angles}),
and the estimate for $\sm$, in the same limit of small $\Fangle$ as
already used above, is $\sm=\Side/2-\r$. This results in the
following:
\begin{align*}
  \Distance(\Side,\r) = \frac{\sqrt3}{2} \Side
  + \left(
    \frac{1}{\p\, C} - \frac{\pi}{6\q\, S}
  \right)\,\left(
    \r\q S + C - 1
  \right) ,
\end{align*}
where $S=\sinh\bigg(\q(\Side/2-\r)\bigg)$ and
$C=\cosh\bigg(\q(\Side/2-\r)\bigg)$,
which is the equation (7) of the main paper.